\documentclass{ws-mpla}

\begin{document}

\markboth{Nannan Wang, Lixin Xu}
{Strong Gravitational Lensing and Its Cosmic Constraints}

\catchline{}{}{}{}{}

\title{Strong Gravitational Lensing and Its Cosmic Constraints}

\author{\footnotesize Nannan Wang, Lixin Xu}

\address{Institute of Theoretical Physics, School of Physics \&
Optoelectronic Technology, Dalian University of Technology, Dalian,
116024, P. R. China\\
lxxu@dlut.edu.cn}

\maketitle

\pub{Received (Day Month Year)}{Revised (Day Month Year)}

\begin{abstract}
In this paper, we propose a new method to use the strong lensing data sets to constrain a cosmological model. By taking the ratio $\mathcal{D}^{obs}_{ij}=\theta_{\mathrm{E_{\mathrm{i}}}}\sigma_{\mathrm{0_{\mathrm{j}}}}^2/\theta_{\mathrm{E_{\mathrm{j}}}}\sigma_{\mathrm{0_{\mathrm{i}}}}^2$ as cosmic observations, one can {\it completely} eliminate the uncertainty caused by the relation $\sigma_{\mathrm{SIS}}=f_{\mathrm{E}}\sigma_0$ which characterizes the relation between the stellar velocity dispersion $\sigma_0$ and  the velocity dispersion $\sigma_{SIS}$. Via our method, a relative tight constraint to the cosmological model space can be obtained, for the spatially flat $\Lambda$CDM model as an example $\Omega_m=0.143_{- 0.143-0.143-0.143}^{+ 0.000769+0.143+0.489}$ in $3\sigma$ regions. And by using this method, one can also probe the nature of dark energy and the spatial curvature of our Universe.

\end{abstract}

\ccode{PACS Nos.: include PACS Nos.}

\section{Introduction}

In 1998, the observations from type Ia supernovae (SN Ia) implied that our Universe is undergoing an accelerated expansion \cite{ref:Riess98,ref:Perlmuter99}. Meanwhile this accelerated expansion obtains independent supports from the anisotropy measurements of Cosmic Microwave Background (CMB)~\cite{Spergel} and the large-scale structure of the universe (LSS)~\cite{Pope}. All these data sets indicate that our Universe is currently dominated by an energy component, dubbed as dark energy, which has a negative pressure and provide a repulsive force to push the Universe into an accelerating expansion phase. The main remained energy component is the cold dark matter which provides an attractive force to form the large scale structure of our Universe. But what is the nature of the dark matter and the dark energy? What are the properties of the dark energy? To understand the nature of this dark side of our Universe, one should use the information from the cosmic observations. 

A very exciting new technology of gravitational lensing provides a strong evidence that when the light rays pass through astronomical objects (galaxies, cluster of galaxies), the cosmological gravitation field bends the paths traveled by light from distant source to us. This fundamental fact carries with it an enormous amount of cosmological promise. The most important is the idea that light paths respond to the distribution of mass. Therefore, the positions of the source and the image are related through the simple lens equation. The information of the invisible matter can be deduced by making a thorough inquiry the relation between sources and images. If one can measure the redshifts of the source and lens, the velocity dispersion of the mass distribution, the separated image, then one might be able to infer the distribution of the mass in our Universe. 

Up to now, the strong gravitational lensing have developed into an important astrophysical tool for probing the cosmology~\cite{Zhu,hong,Chae,H,Mitchell}, the structure formation and the evolution of galaxies ~\cite{Jin,Keeton,Mao,Ofek}. The observations of the source and image combined with lens model can provide the geometric information of our Universe through the ratio between two angular diameter distances $D_{A}(z_l,z_{s})$ (the angular diameter distance between lens and source) and $D_{A}(0,z_{s})$ (the angular diameter distance between the observer and the source). Then one can use the observed data of $D_{A}(z,z_{s})/D_{A}(0,z_{s})$ to constrain the cosmological models, for example please see \cite{Cao} for strong lensing and \cite{Jain} for weak lensing systems. 

Assuming that the gravitational lens can be represented by a singular isothermal sphere (SIS) or singular isothermal ellipsoid (SIE) potential, the Einstein radius in a SIS (or its SIE equivalent) is given as \cite{Ofek,Cao}
\begin{equation}
\theta_{E}=4\pi\frac{D_{A}(z,z_{s})}{D_{A}(0,z_{s})}\frac{\sigma_{SIS}^2}{c^2},\label{eq:ER}
\end{equation}
where $\sigma_{SIS}$ is the velocity dispersion and $c$ is the speed of light. Of course, this assumption is not good for a general lensing system, for example the weak-lensing, but it is a good approximation especially to the strong lensing system where it has $2$ images \cite{Biesiada,Biesiada2011}. The reason is that ellipticity in the lens galaxy mainly affects the relative numbers of two- and four- image lenses but not their overall measures \cite{Keeton1997}. It is clear that if the Einstein radius $\theta_{\mathrm{E}}$ and the velocity dispersion $\sigma_{SIS}$ are measured through lens system, the two angular diameter distances ratio can be derived to constrain the cosmological model. Acyually, the observed Einstein radius $ \theta_{\mathrm{E}}$ can be obtained from Sloan Lens ACS (SLACS) and Lens Structure and Dynamics survey (LSD). 

The big challenge is how to obtain the velocity dispersion $\sigma_{SIS}$. Based on the observations of X-ray, it was argued  that there is a strong indication that dark matter halos are dynamically hotter than the luminous stars ~\cite{White}. Therefore, the velocity dispersion $\sigma_{\mathrm{SIS}}$ of the SIS model is different from the observed stellar velocity dispersion $\sigma_{\mathrm{0}}$. So to fix the problem, one introduces the relation between the stellar velocity dispersion $\sigma_0$ and  the velocity dispersion $\sigma_{SIS}$ in the form of
\begin{equation}
\sigma_{\mathrm{SIS}}=f_{\mathrm{E}}\sigma_0,\label{eq:relation}
\end{equation} 
where $f_{\mathrm{E}}$ is a free parameter \cite{Ofek,Cao}. Then the problem becomes to how to deal with the free parameter $f_{\mathrm{E}}$. It is crucial because it mimics the effects of \cite{Cao}: (i) the systematic rms errors between $\sigma_{\mathrm{0}}$ and $\sigma_{\mathrm{SIS}}$; (ii) the rms error caused by translation the observed image separation into $\theta_{\mathrm{E}}$ in SIS model; (iii) decreasing of the typical image separations due to the softened isothermal sphere potentials \cite{Narayan}. The background matter and the richer environments of early type galaxies can affect the images separation by up to $\pm20\%$ \cite{Martel,Christlein}. It is equivalent to introduce a constant $f_{\mathrm{E}}$ in the range $0.8^{1/2}<f_{\mathrm{E}}<1.2^{1/2}$~\cite{Ofek}. Interestingly, we should mention that the TeVeS/MOND theory of gravity also predicts an isothermal rotation curve in galaxies, and the Einstein ring and velocity dispersions of the CASTLES lenses have been calculated in TeVeS \cite{Zhao2006}. 

In order to eliminate the effects and uncertainties caused by the free parameter $f_{\mathrm{E}}$, in this paper ,we present a new approach that treats $f_{\mathrm{E}}$ as a "nuisance" parameter and harnesses the ratio $\mathcal{D}^{obs}_{ij}=\theta_{\mathrm{E_{\mathrm{i}}}}\sigma_{\mathrm{0_{\mathrm{j}}}}^2/\theta_{\mathrm{E_{\mathrm{j}}}}\sigma_{\mathrm{0_{\mathrm{i}}}}^2$ as cosmic observations to constrain the cosmological model. In this way, the effects and uncertainties caused by the free parameter $f_{\mathrm{E}}$ are eliminated completely. We collected the $66$ strong gravitational lensing systems from SLACS and LSD surveys~\cite{Biesiada,Bolton,Treu,Tonry} as the data points, for the details please see the Table \ref{list}.

\section{Methodology  and data}

Under the assumption of the SIS model \cite{Ofek,Cao}, the Einstein radius $\theta_{E}$ is given in the form of Eq. (\ref{eq:ER}) where $D_{A}$ is the angular diameter distance which is defined as
\begin{equation}
D_A(z;\textbf{p})=\frac{c}{H_{0}} (1+z)\frac{1}{\sqrt{|\Omega_k|}}\text{sinn}\left[\sqrt{|\Omega_k|}\int_{0}^{z} dz'/E(z';\textbf{p}),\right]
\end{equation}
where $\text{sinn}(x)$ is $\sinh(x)$ for $\Omega_k>0$, $\sin(x)$ for $\Omega_k<0$ and $x$ for $\Omega_k=0$ respectively; 
here $ H_{0}$ is the Hubble constant and $E(z;\textbf{p})$ is the dimensionless expansion rate which depends on the redshift $z$ and the cosmological parameters \textbf{p}. In Refs. \cite{White,Gott,Turner,Kochanek}, the authors pointed out that the velocity dispersion $\sigma_{\mathrm{SIS}}$ of the mass distribution and the observation stellar velocity dispersion $\sigma_{\mathrm{0}}$ does not equal but relates via the form (\ref{eq:relation}). In the previous work for example \cite{Cao}, $f_{E}$ was treated as a free model parameter and was found in narrow range of $1$. So, it would be reasonable to assume it is a constant and try to eliminate it completely. Via the relation (\ref{eq:relation}), one can obtain $\mathcal{D}= c^2 \theta_E/4\pi \sigma_{0}^2 f_E^2$, here $\mathcal{D}=D_{A}(z,z_{s})/D_{A}(0,z_{s})$ for different strong lensing system. If one defines the ratio $\mathcal{D}^{obs}_{ij}=\mathcal{D}_{i}/\mathcal{D}_{j}$, where $i,j$ denote the order numbers of the lensing system as marked in Table \ref{list}, 
then one obtains $\mathcal{D}^{obs}_{ij}=\theta_{\mathrm{E_{\mathrm{i}}}}\sigma_{\mathrm{0_{\mathrm{j}}}}^2/\theta_{\mathrm{E_{\mathrm{j}}}}\sigma_{\mathrm{0_{\mathrm{i}}}}^2$ which does not depend on the "nuisance" parameter $f_{E}$ and only relates to the observed data points $\theta_{E}$ and $\sigma_0$ for different strong lensing systems. This is the apparent merit of our methodology. We should notice that this ratio dose not depend on the Hubble parameter. Then it dose not introduce any uncertainty to the results. But on the other side it loses the power to constrain the cosmological models. 

For any strong gravitational lensing system, one can get the theoretical model ratio by calculating
 \begin{equation}
\mathcal{D}^{th}_{ij}(z_d,z_s;\textbf{p})= \mathcal{D}_{i}(z_d,z_s;\textbf{p})/\mathcal{D}_{j}(z_d,z_s;\textbf{p})\label{eq:Dth}.
\end{equation}
Then one can constrain cosmological models by the minimization of $\chi^2$ function which is given by
\begin{equation}
\chi^2(\textbf{p})=
\frac{1}{2\times N}\sum_{i=1}^{N}\sum_{j=1,\neq i}^{N}\frac{(\mathcal{D}_{ij}^{th}(\mathrm{\textbf{p}})-\mathcal{D}_{ij}^{obs})^{2}}{\sigma
_{\mathcal{D},ij}^{2}}, \label{eq:chi2}
\end{equation}
where $N=66$ is the number of the strong lensing system and $\sigma _{\mathcal{D},ij}^{2}$
denotes the $1\sigma$ variance of $\mathcal{D}_{ij}^{obs}$. The $1\sigma$ error $\sigma _{\mathcal{D},ij}^{2}$ of $\mathcal{D}_{ij}^{obs}$ can be calculated via the error propagation equation
\begin{equation}
\sigma_{\mathcal{D},ij}^{2}=4(\mathcal{D}^{obs}_{ij})^{2}(\frac{{\sigma_{\sigma_{{0},j}}}^{2}}{\sigma_{\mathrm{0_{\mathrm{j}}}}^2}+\frac{{\sigma_{\sigma_{{0},i}}}^{2}}{\sigma_{\mathrm{0_{\mathrm{i}}}}^2})\label{eq:errorbar}.
\end{equation}
To obtain the cosmological model parameter space, we use the Markov Chain Monte Carlo (MCMC) method. We modified the publicly available code {\bf cosmoMC} \cite{ref:MCMC} and added one module to calculate the $\chi^2$ from strong lensing system. We ran $8$-chains and stopped sampling when the worst e-values [the variance(mean)/mean(variance) of 1/2 chains] $R-1$ is of the order $0.01$. 

\section{Constraints to cosmological models}

\begin{center}
\begin{figure}[tbh]
\includegraphics[width=14cm]{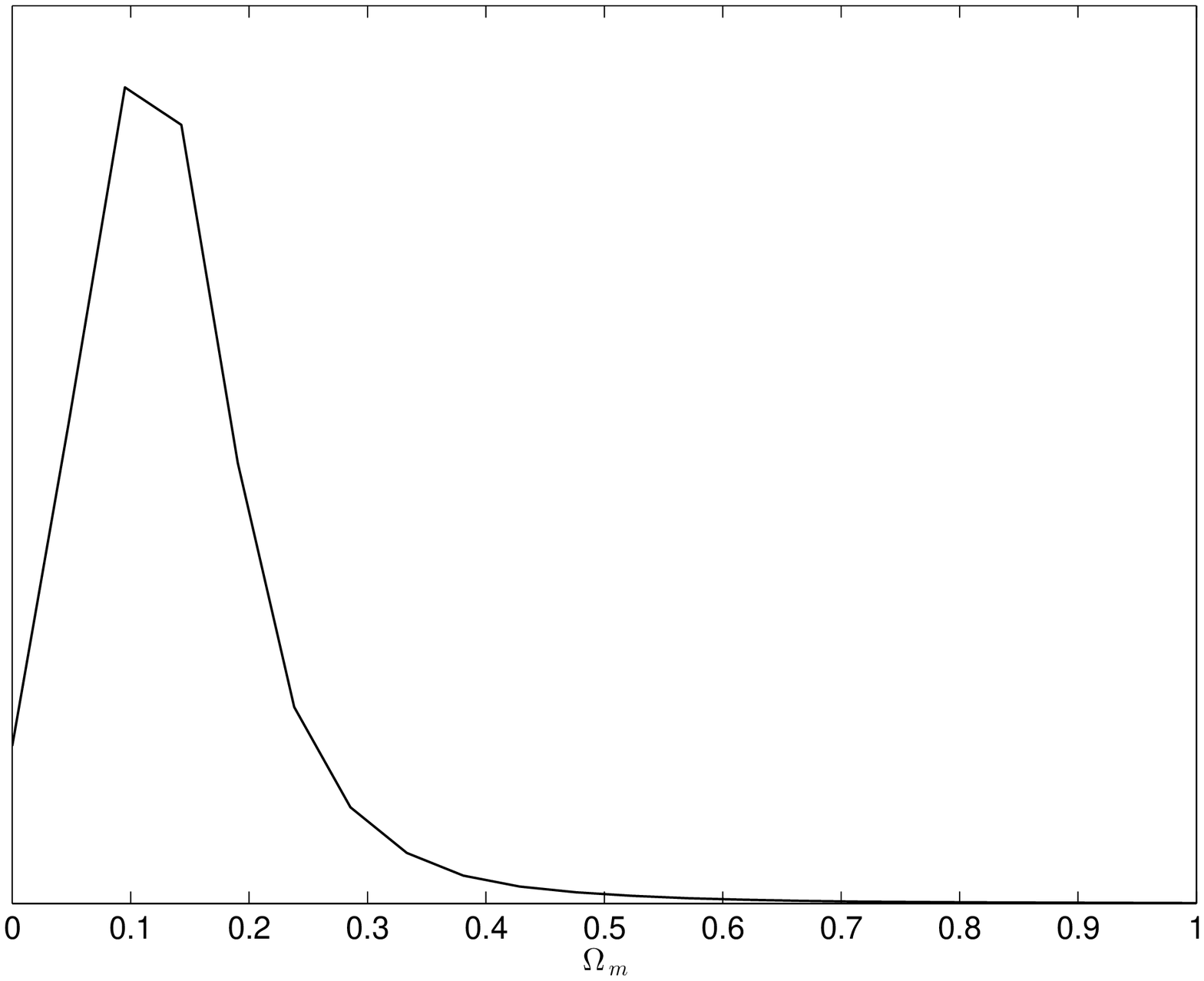}
\includegraphics[width=14cm]{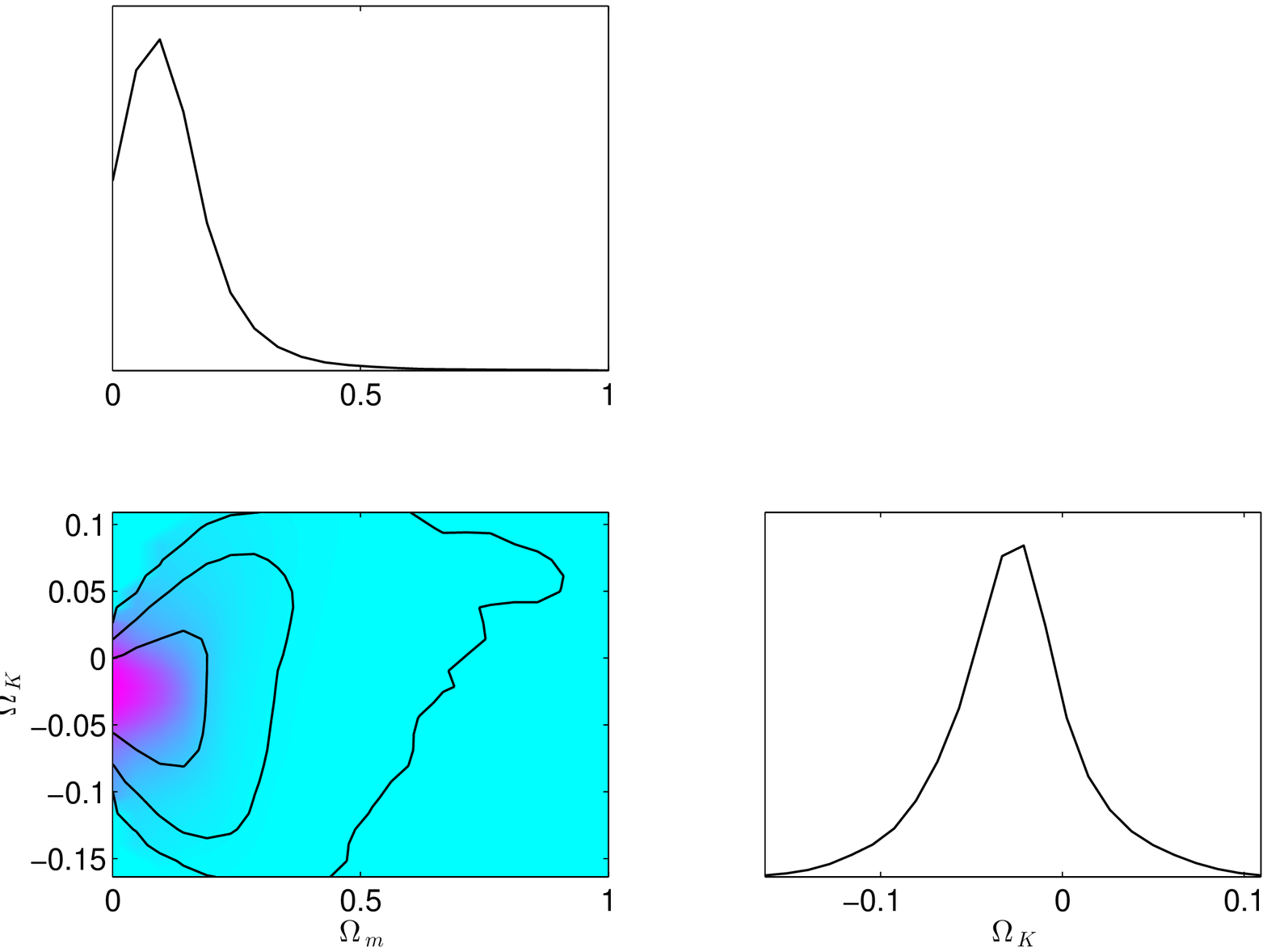}
\caption{The $1-D$ likelihood distribution and $2-D$ contour plot for $\Lambda$CDM model. The up and bottom panels are for the flat and non-flat cases respectively.}\label{fig:lcdm}
\end{figure}
\end{center}

In the simplest $\Lambda$CDM model, the dark energy is a cosmological constant $\Lambda$ which has 
a constant equation of state (EoS) $w_{\Lambda}=p_{\Lambda}/\rho_{\Lambda}=-1$. For this simple cosmological model, the Hubble parameter free expansion rate is written as 
\begin{equation}
E^2(z;\textbf{p})=\Omega_m(1+z)^3 +\Omega_k(1+z)^2 +\Omega_\Lambda,
\end{equation}
where $\Omega_m$, $\Omega_k$ and $\Omega_\Lambda=1-\Omega_m-\Omega_k$ are the dimensionless energy density for the cold dark matter, spatial curvature and the cosmological constant respectively. In this model, we will consider the spatially flat and non-flat cases respectively. For the spatially flat case, it has only one free model parameter: $\textbf{p}=\{\Omega_{m}\}$. For the non-flat case, it has the model parameter: $\textbf{p}=\{\Omega_{m},\Omega_k\}$. The results are shown in the Table \ref{tab:resultslcdm} and the Figure \ref{fig:lcdm}.

\begin{table}[h]
\tbl{The model parameter space with errors for $\Lambda$CDM model with and without spatial curvature.}
{\begin{tabular}{@{}ccc@{}} \toprule
Parameters & flat & non-flat \\
\colrule
$\Omega_m$ & $    0.143_{-    0.143-    0.143-    0.143}^{+    0.000769+    0.143+    0.489}$ & $    0.130_{-    0.130-    0.130-    0.130}^{+    0.00740+    0.169+    0.597}$\\
$\Omega_k$ & $-$  & $   -0.0278_{-    0.0318-    0.0785-    0.120}^{+    0.0303+    0.0803+    0.122}$\\
\botrule
\end{tabular}\label{tab:resultslcdm} }
\end{table}

\begin{center}
\begin{figure}[tbh]
\includegraphics[width=14cm]{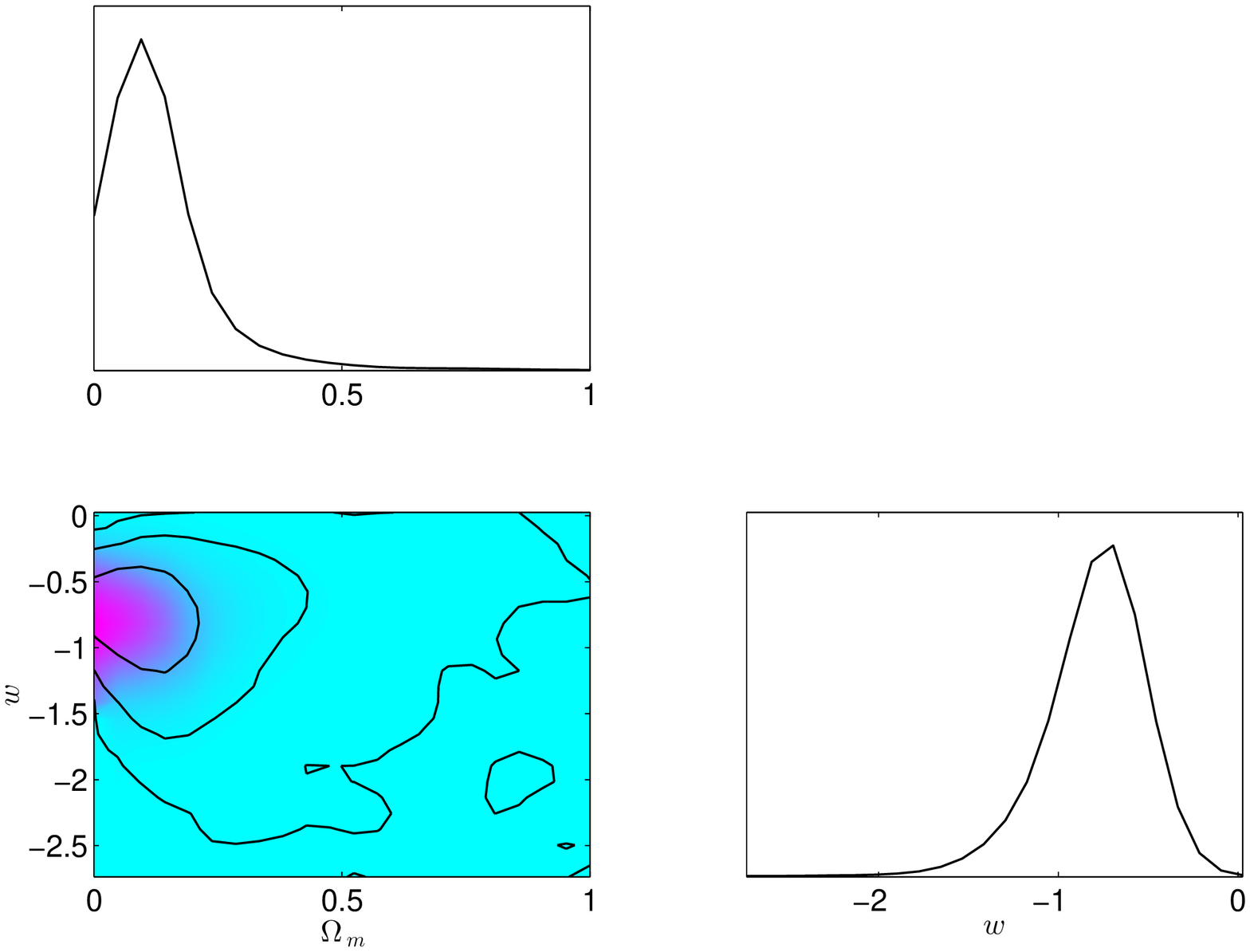}
\includegraphics[width=14cm]{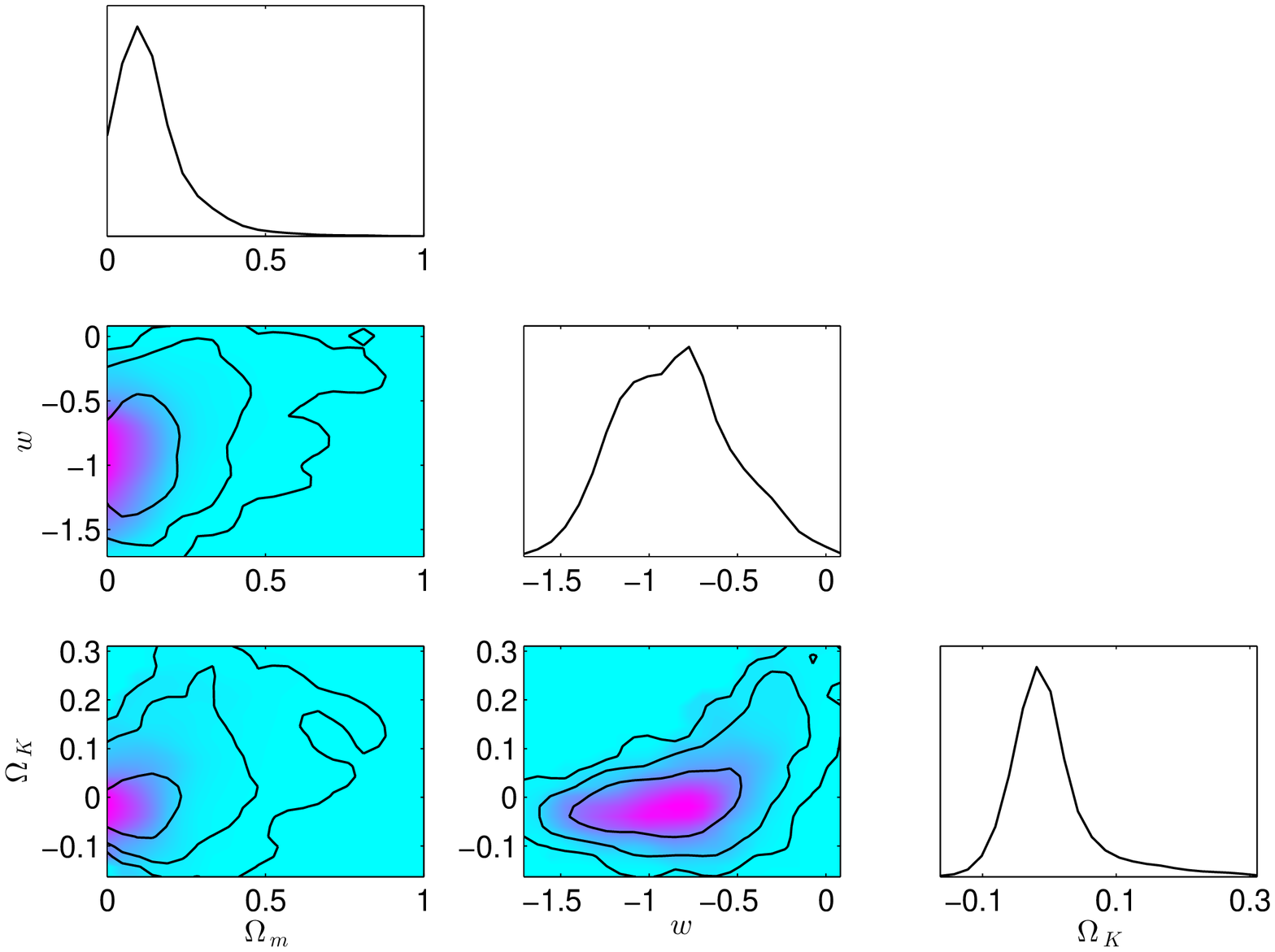}
\caption{The $1-D$ likelihood and $2-D$ contour plots for $w$CDM model. The up and bottom panels are for the flat and non-flat cases respectively.}\label{fig:lwmb}
\end{figure}
\end{center}

Allowing for a deviation from the simple $w_{\Lambda}=-1$ case, for the dark energy with a constant EoS $w$, one has the Hubble free expansion rate in the following form
\begin{equation}
E^2(z;\textbf{p})= \Omega_m(1+z)^3 +\Omega_k(1+z)^2 + \Omega_x(1+z)^{3(1+w)},
\end{equation}
where $\Omega_x=1-\Omega_m-\Omega_k$. It has two free model parameters $\textbf{p}=\{\Omega_m,w\}$ for the flat case and three model parameters $\textbf{p}=\{\Omega_m,\Omega_k, w\}$ for the non-flat case. The corresponding results are shown in the Table \ref{tab:resultswcdm} and the Figure \ref{fig:lwmb}.

\begin{table}[h]
\tbl{The model parameter space with errors for $w$CDM model with and without spatial curvature.}
{\begin{tabular}{@{}ccc@{}} \toprule
Parameters & flat & non-flat \\
$\Omega_m$ & $    0.141_{-    0.141-    0.141-    0.141}^{+    0.00269+    0.195+    0.677}$ & $    0.150_{-    0.150-    0.150-    0.150}^{+    0.00838+    0.206+    0.581}$ \\
$w$ & $   -0.802_{-    0.235-    0.650-    1.673}^{+    0.243+    0.443+    0.696}$ & $   -0.856_{-    0.321-    0.557-    0.775}^{+    0.322+    0.630+    0.859}$\\
$\Omega_k$ & $-$ & $    0.00132_{-    0.0471-    0.0871-    0.134}^{+    0.0412+    0.189+    0.289}$ \\
\botrule
\end{tabular}\label{tab:resultswcdm}}
\end{table}

By comparison to the results as shown in Ref. \cite{Cao}, our method can give a relative tight constraint to the cosmological model parameter space. But the effects caused by $f_E$ are removed completely. And, via our method, the spatial curvature can also be constrained. To compare our results to that from SN Ia Union2.1 \cite{ref:Union21}, we reran the code and found the corresponding results: $\Omega_m=0.296_{-  0.0232-    0.0690}^{+    0.0195+    0.0778}$ for the flat $\Lambda$CDM model and $\Omega_m= 0.258_{-    0.258-    0.258}^{+    0.0632+    0.174}$ and $w=-0.973_{-    0.305-    0.737}^{+    0.303+    0.409}$ for the flat $w$CDM model respectively. Clearly, our results are compatible with that from SN Ia. But the strong lensing system favor lower values of $\Omega_m$.

\section{Conclusion} In this paper, we present a new method to use the strong lensing system to constrain the cosmological models. Assuming that the gravitation lens can be represented by a singular isothermal sphere (SIS) or singular isothermal ellipsoid ( SIE) potential, we eliminated completely the uncertainty caused by $f_{E}$ which characterizes the relation between the stellar velocity dispersion $\sigma_0$ and  the velocity dispersion $\sigma_{SIS}$. By taking the ratios of $\mathcal{D}^{obs}_{ij}=\theta_{\mathrm{E_{\mathrm{i}}}}\sigma_{\mathrm{0_{\mathrm{j}}}}^2/\theta_{\mathrm{E_{\mathrm{j}}}}\sigma_{\mathrm{0_{\mathrm{i}}}}^2$ as cosmic observations, one can constrain the cosmological model parameter space very well. And it can be used to probe the nature of dark energy and the spatial curvature of our Universe.

\appendix

\section{Strong lensing data Points}

\begin{table}[h]
\tbl{Values of $\mathcal{D}^{obs}=\theta_{\mathrm{E_{\mathrm{i}}}}\sigma_{\mathrm{0_{\mathrm{j}}}}^2/\theta_{\mathrm{E_{\mathrm{j}}}}\sigma_{\mathrm{0_{\mathrm{i}}}}^2$ from lensing SLACS+LSD lens samples.}
{\begin{tabular}{@{}lccccccc@{}} \toprule
System & z$_l$ & z$_s$  & $\theta_E ['']$ & $\sigma_0\;[km/s]$ & $\mathcal{D}^{obs}$ & $\sigma_\mathcal{D}$ & ref \\
SDSS J0037-0942 & 0.1955 & 0.6322&1.47&$282\pm11$  & 0.7931  & 0.1693 & ~\cite{Biesiada,Grillo}\\
SDSS J0216-0813 & 0.3317 & 0.5235 & 1.15 & $ 349 \pm 24$ & 1.5527  & 0.3753 & ~\cite{Biesiada,Grillo}\\
SDSS J0737+3216 & 0.3223 & 0.5812 & 1.03 & $ 326 \pm 16$ & 1.5126  & 0.3753 & ~\cite{Biesiada,Grillo}\\
SDSS J0912+0029 & 0.1642 & 0.324& 1.61 & $ 325 \pm 12$  & 0.9618  & 0.2039 & ~\cite{Biesiada,Grillo}\\
SDSS J0956+5100 & 0.2405 & 0.47   & 1.32 & $ 318 \pm 17$  & 1.1231  & 0.2535 & ~\cite{Biesiada,Grillo}\\
SDSS J0959+0410 & 0.126  & 0.5349& 1.00 & $ 229 \pm 13$  & 0.7688  &0.2172 & ~\cite{Biesiada,Grillo}\\
SDSS J1250+0523 & 0.2318 & 0.795  & 1.15 & $ 274 \pm 15$  & 0.9571  & 0.0996 & ~\cite{Biesiada,Grillo}\\
SDSS J1330-0148 & 0.0808 & 0.7115 & 0.85 & $ 195 \pm 10$  & 0.6558 & 0.1467 & ~\cite{Biesiada,Grillo}\\
SDSS J1402+6321 & 0.2046 & 0.4814  & 1.39 & $ 290 \pm 16$ & 0.887 & 0.2016 & ~\cite{Biesiada,Grillo}\\
SDSS J1420+6019 & 0.0629 & 0.5352 & 1.04 & $ 206 \pm 5$ &0.5982 & 0.1224 & ~\cite{Biesiada,Grillo}\\
SDSS J1627-0053 & 0.2076 & 0.5241& 1.21 & $ 295 \pm 13$  & 1.0544 & 0.2292 &~ \cite{Biesiada,Grillo}\\
SDSS J1630+4520 & 0.2479 & 0.7933  & 1.81 & $ 279 \pm 17$ & 0.6305 & 0.147& ~\cite{Biesiada,Grillo}\\
SDSS J2300+0022 & 0.2285 & 0.4635 & 1.25 & $ 305 \pm 19$ & 1.091 & 0.256 & ~\cite{Biesiada,Grillo}\\
SDSS J2303+1422 & 0.1553 & 0.517 & 1.64 & $ 271 \pm 16$ & 0.6565 & 0.1518& ~\cite{Biesiada,Grillo}\\
SDSS J2321-0939 & 0.0819 & 0.5324 & 1.57 & $ 245 \pm 7$ & 0.5605&0.116 & ~\cite{Biesiada,Grillo}\\
Q0047-2808 & 0.485 & 3.595 & 1.34 & $ 229 \pm 15$ &0.5737 & 0.1366 & ~\cite{Biesiada,Grillo}\\
CFRS03-1077 & 0.938 & 2.941 & 1.24 & $ 251 \pm 19$  & 0.7448& 0.1861& ~\cite{Biesiada,Grillo}\\
HST 14176 & 0.81 & 3.399 & 1.41 & $ 224 \pm 15$ & 0.5217 &  0.1250& ~\cite{Biesiada,Grillo}\\
HST 15433 & 0.497 & 2.092  & 0.36 & $ 116 \pm 10$  & 0.548 & 0.1442 & ~\cite{Biesiada,Grillo}\\
MG 2016 & 1.004 & 3.263 & 1.56 & $ 328 \pm 32$ & 1.0110 & 0.2816 & ~\cite{Biesiada,Grillo}\\
SDSS J0029-0055  & 0.227  & 0.9313  &0.96 & $ 229 \pm 18$  &  0.8008 &  0.2029  &~\cite{Bolton,Gobat}\\
SDSS J0044+0113  &   0.1196 &   0.1965 &0.79  & $ 266 \pm 13$   &   1.3130  &  0.2908 &~\cite{Bolton,Gobat}\\
SDSS J0109+1500  &   0.2939 &   0.5248 &0.69  & $ 251 \pm 19$  &1.3386 & 0.3344  &\cite{Bolton,Gobat}\\
SDSS J0252+0039  &   0.2803 &   0.9818 &1.04 & $ 164 \pm 12$   &   0.3791 &  0.0936  &\cite{Bolton,Gobat}\\
SDSS J0330-0020  &   0.3507  &   1.0709&1.1  & $ 212 \pm 21$   &  0.599  &  0.1681 &\cite{Bolton,Gobat}\\
SDSS J0405-0455  &   0.0753 &   0.8098 &0.8   & $ 160 \pm 8$  &  0.4691 &   0.1044  &\cite{Bolton,Gobat}\\
SDSS J0728+3835  &   0.2058 &   0.6877 &1.25  & $ 214 \pm 11$   &  0.5371  &   0.1202  &\cite{Bolton,Gobat}\\
SDSS J0822+2652  &   0.2414 & 0.5941&1.17  & $ 259 \pm 15$    & 0.8405 &  0.1934  &\cite{Bolton,Gobat}\\
SDSS J0841+3824  &   0.1159 &   0.6567 &1.41   & $ 225 \pm 11$  &  0.5264 &  0.1166  &\cite{Bolton,Gobat}\\
SDSS J0935-0003  &   0.3475 &   0.467 &0.87  & $ 396 \pm 35$    &  2.6424  &  0.7029 &\cite{Bolton,Gobat}\\
SDSS J0936+0913  &   0.1897 &   0.588 &1.09  & $ 243 \pm 12$    &  0.7942 & 	0.1763  &\cite{Bolton,Gobat}\\
SDSS J0946+1006  &   0.2219 &   0.6085 &1.38  & $ 263 \pm 21$   &  0.7348  &  0.1873 &\cite{Bolton,Gobat}\\
SDSS J0955+0101 &   0.1109 &   0.3159 &0.91   & $ 192\pm 13$  &   0.5939  &   	0.1428  &\cite{Bolton,Gobat}\\
SDSS J0959+4416 &   0.2369 &   0.5315&0.96   & $ 244 \pm 19$   &  0.9092 &  0.2296 &\cite{Bolton,Gobat}\\
SDSS J1016+3859  &   0.1679 &   0.4394 &1.09  & $ 247\pm 13$   &   0.8205 &  0.1845  &\cite{Bolton,Gobat}\\
SDSS J1020+1122  &   0.2822 &   0.553&1.2   & $ 282 \pm 18$    &   0.9715  &  0.2295  &\cite{Bolton,Gobat}\\
SDSS J1023+4230  &   0.1912 &   0.696 &1.41   & $ 242 \pm 15$   &  0.6089  &  0.1426  &\cite{Bolton,Gobat}\\
SDSS J1029+0420  &   0.1045 &   0.6154 &1.01  & $ 210 \pm 11$   &  0.6401 &  0.1438  &\cite{Bolton,Gobat}\\
SDSS J1032+5322  &   0.1334 &   0.329  &1.03  & $ 296 \pm 15$   &   1.2470 &  0.2782  &\cite{Bolton,Gobat}\\
SDSS J1103+5322  &   0.1582 &   0.7353 &1.02   & $ 196 \pm 12$  &  0.5521 & 0.1289 &\cite{Bolton,Gobat}\\
SDSS J1106+5228  &   0.0955 &   0.4069 &1.23  & $ 262\pm 13$   &  0.8181 &  0.1818 &\cite{Bolton,Gobat}\\
SDSS J1112+0826  &   0.273 &   0.6295 &1.49  & $ 320 \pm 20$    &   1.0075 &  	0.2366  &\cite{Bolton,Gobat}\\
SDSS J1134+6027  &   0.1528 &   0.4742 &1.1  & $ 239 \pm 12$   &   0.7613 &  0.167 &\cite{Bolton,Gobat}\\
SDSS J1142+1001  &   0.2218 &   0.5039  &0.98  & $ 221 \pm 22$  &   0.7306 & 0.2055  &\cite{Bolton,Gobat}\\
SDSS J1143-0144  &   0.106 &   0.4019 &1.68   & $ 269 \pm 13$   &  0.6314 &  	0.14 &\cite{Bolton,Gobat}\\
SDSS J1153+4612  &   0.1797  &   0.8751&1.05  & $ 226 \pm 15$   &  0.7131 &  0.1704 &\cite{Bolton,Gobat}\\
SDSS J1204+0358  &   0.1644 &   0.6307&1.31  & $ 267 \pm 17$    &   0.7978 &  0.1883  &\cite{Bolton,Gobat}\\
SDSS J1205+4910  &   0.215 &   0.4808 &1.22   & $ 281 \pm 14$   &  0.9488 &   0.211 &\cite{Bolton,Gobat}\\
SDSS J1213+6708  &   0.1229  &   0.6402 &1.42  & $ 292 \pm 15$  &   0.8803&  0.1969 &\cite{Bolton,Gobat}\\
SDSS J1218+0830  &   0.135 &   0.7172 &1.45  & $ 219 \pm 11$    &  0.4849 &   0.108 &\cite{Bolton,Gobat}\\
SDSS J1403+0006  &   0.1888 &   0.473 &0.83    & $ 213 \pm 17$  &  0.8013 &  0.2043 &\cite{Bolton,Gobat}\\
SDSS J1416+5136  &   0.2987  &   0.8111&1.37  & $ 240 \pm 25$   &  0.6164  &   	0.1775 &\cite{Bolton,Gobat}\\
SDSS J1430+4105  &   0.285 &   0.5753  &1.52  & $ 322 \pm 32$   &   1   &   0.2811  &\cite{Bolton,Gobat}\\
SDSS J1432+6317  &   0.123 &   0.6643&1.26   & $ 199 \pm 10$    &  0.4608 & 0.1026  &\cite{Bolton,Gobat}\\
SDSS J1436-0000  &   0.2852 &   0.8049 &1.12  & $ 224 \pm 17$   &  0.6568  &  0.1642 &\cite{Bolton,Gobat}\\
\botrule
\end{tabular}\label{list}}
\end{table}

\begin{table}[h]
\tbl{Continue: Values of $\mathcal{D}^{obs}=\theta_{\mathrm{E_{\mathrm{i}}}}\sigma_{\mathrm{0_{\mathrm{j}}}}^2/\theta_{\mathrm{E_{\mathrm{j}}}}\sigma_{\mathrm{0_{\mathrm{i}}}}^2$ from lensing SLACS+LSD lens samples.}
{\begin{tabular}{@{}lccccccc@{}} \toprule
System & z$_l$ & z$_s$  & $\theta_E ['']$ & $\sigma_0\;[km/s]$ & $\mathcal{D}^{obs}$ & $\sigma_\mathcal{D}$ & ref \\
SDSS J1443+0304  &   0.1338 &   0.4187&0.81   & $ 209 \pm 11$   &   0.7906 &  0.1778 &\cite{Bolton,Gobat}\\
SDSS J1451-0239  &   0.1254 &   0.5203 &1.04   & $ 223 \pm14 $  &  0.701 &   0.1648 &\cite{Bolton,Gobat}\\
SDSS J1525+3327  &   0.3583 &   0.7173 &1.31  & $ 264 \pm 26$   &   0.78 &   0.2183 &~\cite{Bolton,Gobat}\\
SDSS J1531-0105  &   0.1596 &   0.7439 &1.71   & $ 279 \pm 14$  &   0.6673 &   0.1486 &~\cite{Bolton,Gobat}\\
SDSS J1538+5817  &   0.1428 &   0.5312&1   & $ 189 \pm 12$   &   0.5237   &   	0.1235 &~\cite{Bolton,Gobat}\\
SDSS J1621+3931  &   0.2449  &   0.6021 &1.29  & $ 236 \pm 20$  &  0.6329 &   0.1653 &~\cite{Bolton,Gobat}\\
SDSS J1636+4707  &   0.2282 &   0.6745  &1.09  & $ 231 \pm 15$  &  0.7177 &   0.1704 &~\cite{Bolton,Gobat}\\
SDSS J2238-0754 &   0.1371  &   0.7126 &1.27  & $ 198 \pm 11$  &   0.4525  &   0.1030  &~\cite{Bolton,Gobat}\\
SDSS J2341+0000  &   0.186 &   0.807 &1.44 & $ 207\pm 13$      &   0.4362  &  0.1026 &~\cite{Bolton,Gobat}\\
MG1549+3047 &   0.11 &   1.17  &1.15    & $ 227 \pm 18$    &   0.6569   &  0.167  &~\cite{Treu,Lehar}\\
PG1115+080  &   0.31  &   1.72  &1.21    & $ 281 \pm 25$   &   0.9567 &  0.2552&~\cite{Tonry}\\
\botrule
\end{tabular}}
\end{table}


\section*{Acknowledgements} The authors thank an anonymous referee for helpful improvement of this paper. L. Xu's work is supported in part by NSFC under the Grants No. 11275035 and "the Fundamental Research Funds for the Central Universities" under the Grants No. DUT13LK01.


\begin{thebibliography}{0}

\bibitem{ref:Riess98} A.G. Riess, {\it et al.}, Astron. J. 116, 1009(1998) [astro-ph/9805201]. 
\bibitem{ref:Perlmuter99} S. Perlmutter, {\it et al.}, Astrophys. J. {\bf 517}, 565(1999) [astro-ph/9812133].
\bibitem{Spergel} D.N.Spergel \textit{et al.}, Astrophys. J. Suppl. Ser \textbf{148}, 175 (2003).
\bibitem{Pope} A. C. Pope \textit{et al.}, Astrophys. J. \textbf{607}, 65 (2004).
\bibitem{Zhu} Z. H. Zhu, Mod. Phys. Lett. \textbf{A 15}, 1023 (2000) [astor-ph/0010351].
\bibitem{hong} Z. H. Zhu, Int. J. Mod. Phys. \textbf{D 9}, 591 (2000) [astor-ph/0107233].
\bibitem{Chae} K. H. Chae, Mon. Not. Roy. Asrron. Soc. \textbf{346}, 746 (2003).
\bibitem{H} K. H. Chae \textit{et al.}, Astrophys. J. \textbf{607}, L71 (2004) [astor-ph/0403256].
\bibitem{Mitchell} J. L. Mitchell \textit{et al.}, Astrophys. J. \textbf{622}, 81 (2005) [astor-ph/0401138].
\bibitem{Jin} K. J. Jin, Y. Z. Zhang and Z. H. Zhu,  Phys. Lett. \textbf{A 264}, 335 (2000) [gr-qc/9907035].
\bibitem{Keeton} C. R. Keeton, Astrophys. J. \textbf{561}, 46 (2001).
\bibitem{Mao} S.D. Mao and P. Schneider, Mon. Not. Roy. Asrron. Soc. \textbf{295}, 587 (1998) [astor-ph/9707187].
\bibitem{Ofek} E.O. Ofek, H. W. Rix and D. Maoz, Mon. Not. Roy. Asrron. Soc. \textbf{343}, 639 (2003) [astro-ph/0305201].
\bibitem{Cao} S. Cao and Z. H. Zhu, JCAP, \textbf{03}, 016 (2012) [astor-ph/1105.6226].
\bibitem{Biesiada} M. Biesiada, A. Piorkowska and B. Malec, Mon. Not. Roy. Astron. Soc. \textbf{406} 1055 (2010) [arXiv:1105.0946].
\bibitem{Biesiada2011} M. Biesiada, B. Malec and A. Piorkowska, Res. Astron. Astrophys. 11, 641(2011). 
\bibitem{Keeton1997} C. R. Keeton, C. S. Kochanek, U. Seljak, Astrophys. J. {\bf 482}, 604(1997).
\bibitem{Jain} B. Jain, A. Taylor, Phys. Rev. Lett. \textbf{91}, 141302(2003).
\bibitem{White} R. E. White and D. S. Davis, Bull. Am. Astron. Soc. \textbf{28}, 1323 (1998).
\bibitem{Narayan} R. Narayan and M. Bartelmann, [astor-ph/9606001].
\bibitem{Martel} H. Martel, P. Premadi and R. Matzner, Astrophys. J. \textbf{570}, 17 (2002).
\bibitem{Christlein} D. Christlein,  Astrophys. J. \textbf{545}, 145 (2000).
\bibitem{ref:MCMC} http://cosmologist.info/cosmomc/; A. Lewis and S. Bridle, Phys. Rev. D 66, 103511 (2002).
\bibitem{Zhao2006} H. Zhao, D. J. Bacon, A. N. Taylor, K. Horne, Mont.Not. Roy. Astro. Soc., 368,171(2006). 
\bibitem{Gott} J. R. Gott, Ann. Rev. Astr. Ap.,\textbf{15}, 239(1977).
\bibitem{Turner} E. L. Turne, J. P. Ostriker, J. R. GoTT , Astrophys. J. \textbf{284}, (1984).
\bibitem{Kochanek} C. S. Kochanek, Astrophys. J. \textbf{384}, (1992).
\bibitem{Grillo} C. Grillo, M. Lombardi and G. Bertin, Astron. Astrophys. \textbf{477}, 397 (2008) 397.
\bibitem{Bolton} A. S. Bolton \textit{et al.}, Astrophys. J. \textbf{682}, 964 (2008) [arXiv:0805.1931].
\bibitem{Gobat} C. Grillo, R.Gobat, M. Lombardi and P. Rosati, [arXiv:astor-ph/0904.3282].
\bibitem{Treu} T. Treu and L. V. E. Koopmans, Mon. Not. R. Astron. Soc. \textbf{343}, L29 (2003) [arXiv:astor-ph/0306045].
\bibitem{Lehar}  J. Lehar, G. I. Langston, A. Silber, C. R. Lawrence and B. F. Burke, Astrophys. J. \textbf{105}, 847 (1993)
\bibitem{Tonry} J. L. Tonry, Astron. J. \textbf{115}, 1 (1998) .
\bibitem{ref:Union21} N. Suzuki, et al. (Supernova Cosmology Project
Collaboration), arXiv:1105.3470 [astro-ph.CO], http://supernova.lbl.gov/Union/.
 
\end{thebibliography}
\end{document}